# Full-Field Brillouin Microscopy with a Scanning Fabry–Perot Interferometer

*Mikolaj Pochylski*

Faculty of Physics and Astronomy, Adam Mickiewicz University, Poznań, Poland

Email: pochyl@amu.edu.pl

## Abstract

Brillouin microscopy is an emerging optical technique for probing mechanical properties with submicron resolution, offering fully non-contact, label-free operation. Despite its unique capabilities, broader adoption has been limited by slow acquisition speeds, particularly in systems based on scanning Fabry–Perot interferometers (FPIs). Based on prior implementations, FPIs have typically been considered too slow for practical imaging, particularly when both spatial and spectral precision are required. Here, we demonstrate that a standard multi-pass tandem FPI can be repurposed for full-field Brillouin imaging when operated in a spectral filtering mode. Combined with light-sheet illumination for uniform, low-dose excitation, this configuration enables rapid, spatially resolved acquisition of Brillouin spectra. By restricting scanning to a narrow frequency range around the Brillouin peak, we acquired a full 2D image within one minute, achieving millisecond-scale single pixel dwell times and micrometer-scale spatial resolution. The system uniquely supports Brillouin emission imaging at selected frequency shifts, a capability not available with other spontaneous Brillouin implementations. Results from synthetic and biological specimens demonstrate how existing FPI-based setups can be extended to full-field imaging and outline a pathway toward future dedicated FPI instruments optimized for high-speed, high-contrast Brillouin microscopy.

## 1. Introduction

Brillouin microscopy is rapidly emerging as a powerful technique for non-invasive, label-free, and three-dimensional imaging of mechanical properties in biological systems [1–4]. By probing GHz-frequency viscoelastic moduli through light scattering from thermally excited acoustic phonons, Brillouin microscopy enables direct mapping of viscoelastic parameters in living cells and tissues at submicron resolution, all without physical contact. Despite its unique advantages, Brillouin microscopy has traditionally been limited by slow acquisition speeds. These limitations arise from both fundamental and methodological factors: fundamentally, the spontaneous Brillouin process has an inherently low scattering cross-section, producing weak signals that require long integration times; methodologically, most systems rely on sequential spatial and spectral scanning. This is particularly true for setups based on scanning Fabry–Perot interferometers (FPIs). To address these challenges, two main strategies have emerged: one enhances the signal through stimulated Brillouin scattering [5,6] using nonlinear optical gain, while the other improves acquisition efficiency by applying spectral and spatial multiplexing to spontaneous Brillouin scattering, most commonly with VIPA-based interferometer [7–12]. Line-scanning with VIPA has significantly boosted imaging throughput by enabling one-dimensional spatial multiplexing. A natural next step is to expand this concept to full two-dimensional spatial multiplexing. However, VIPA technology does not support this mode of operation, prompting the exploration of alternative solutions. One such approach employed an atomic gas absorption cell as a narrow-band scanning monochromator. Although full-field imaging was demonstrated [13], and its potential for Brillouin spectroscopy explored [14], this concept has yet to result in a fully functional full-field Brillouin imaging system. Only recently has Fourier-transform Brillouin imaging, based on a scanning Michelson-type interferometer [15], achieved true full-field acquisition in the spontaneous regime, confirming anticipated gains in spectrum acquisition speed. However, the limited spectral contrast of both of these full-field approaches (similar to that of VIPA-based spectrometers) necessitates additional suppression of the elastic background, typically through gas notch filters, to effectively isolate the weak Brillouin peak.

On the other hand, scanning multi-pass Fabry–Perot interferometers (FPIs) are well known for their exceptional spectral resolution and contrast, and for their ability to efficiently transmit light at selected frequencies. While FPIs were widely used for single-point Brillouin imaging [16–20] and have also been successfully applied in full-field imaging across other disciplines [21–27], their potential for full-field Brillouin microscopy has remained unexplored. This may be due, in part, to the particularly demanding spectral requirements of Brillouin imaging, significantly more stringent than in previous applications. In Brillouin spectroscopy, the weak inelastic signal lies spectrally close to the intense elastic feature, requiring extremely high spectral contrast to isolate weak Brillouin signature. Additionally, comparisons between conventional implementations of FPI with frequency-multiplexed interferometer (VIPA) systems [28,29] may have contributed to generalized view that frequency-scanning FPIs may be inherently too slow for practical imaging, particularly when both high spatial and spectral resolution are required.

In this work, we demonstrate that a multi-pass tandem FPI can be operated in full-field spectral filtering mode to achieve Brillouin maps at practical acquisition rates. The overall speed-up arises from two key factors: (1) two-dimensional spatial multiplexing, made possible by the FPI's ability to transmit light from extended sources; and (2) the spectral localization of the Brillouin line, which allows scanning to be confined to a narrow frequency range around the peak. Combined with light-sheet illumination—which ensures uniform excitation across the field of view, provides optical sectioning, and minimizes photodose—this configuration offers a simple and efficient route to high-resolution mechanical imaging.

## 2. Results

### 2.1 Full-field Brillouin imaging using scanning FPI.

In our implementation, a canonical Sandercock-type multi-pass tandem Fabry–Perot interferometer (TFPI) was adapted for imaging by redirecting the transmitted light to a CCD camera and operating the system through a custom-designed control interface (see Supplementary Section 1). In this canonical design, light traverses two slightly detuned FPI cavities in a six-pass configuration, resulting in selective transmission of a single interference order. This configuration yields exceptional spectral performance, with elastic signal suppression exceeding 150 dB, crucial for isolating the weak Brillouin signal in the presence of strong elastic scattering.

A schematic of our full-field imaging system is shown in Fig. 1a. To enable spatially resolved Brillouin imaging, the tandem Fabry–Perot interferometer (TFPI) is combined with light-sheet illumination, which provides optical sectioning, uniform excitation across the field of view, and minimal light exposure to the sample. Brillouin-scattered light is collected orthogonally and passed through the interferometer, which is tuned to transmit a narrow spectral band at a fixed mirror separation. The filtered light is then projected as a two-dimensional spectral image onto a CCD camera. A stack of such sequentially acquired images, each corresponding to a discrete frequency, is used to reconstruct the Brillouin spectrum at every pixel, enabling spatially resolved mechanical maps to be derived through model-based fitting.

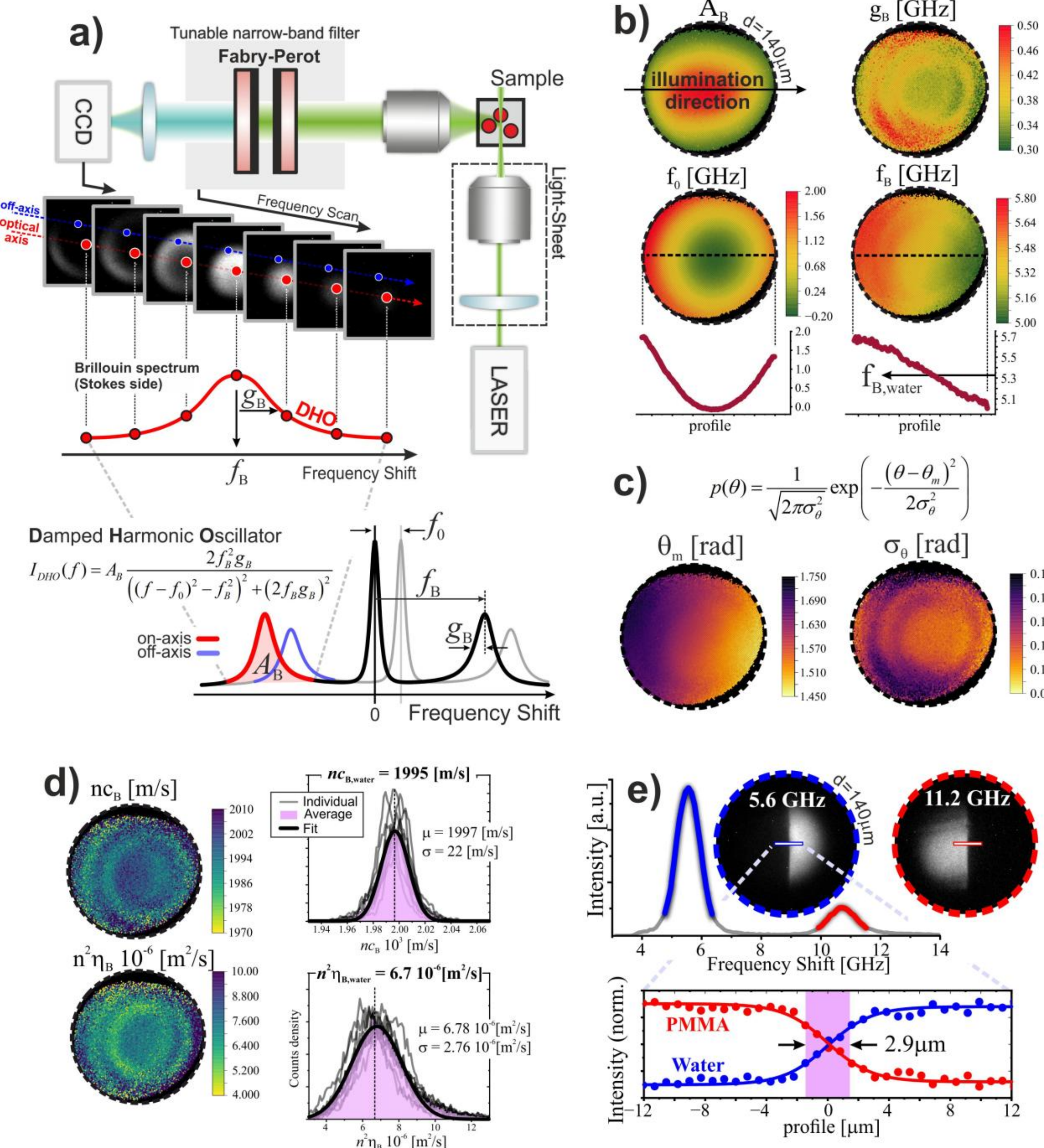

**Fig. 1. Principle and performance of full-field Brillouin imaging using a scanning Fabry–Perot interferometer. a**, A light sheet illuminates the sample, and orthogonally scattered light passes through a scanning Fabry–Perot interferometer

(FPI), acting as a narrowband spectral filter. For each mirror separation, single spectral image is captured. A stack of such images reconstructs the Brillouin spectrum at every pixel (as schematically illustrated), enabling extraction of mechanical parameters through model fitting, such as with the damped harmonic oscillator (DHO) model. **b**, DHO parameter maps (amplitude $A_B$, shift $f_B$, linewidth $g_B$, and offset $f_0$) from water show strong field-dependence due to scattering angle filtering and FPI optics misalignments. The dashed circles mark the field of view, limited by the physical aperture of the TFPI entrance pinhole. The $f_B$ map forms an inclined plane along the illumination axis with the expected water value (~5.3 GHz) occurring only at the FPI optical axis, while $f_0$ shows a parabolic profile, indicating spectral blue-shifts at off-axis pixels. **c**, Calibration maps of the Gaussian scattering angle distribution function, $p(\theta)$. The mean scattering angle $\theta_m$ and the width of distribution (standard deviation) $\sigma_\theta$, were obtained from fitting Eq. 5 to water spectra with fixed values of reduced mechanical properties. **d**, Corrected maps of reduced mechanical parameters in water: $nc_B$ and $n^2\eta_B$. Histograms of the retrieved mechanical parameters were compiled from ten independent measurements (gray lines). The averaged histogram (shaded area) for each parameter was fitted with a normal distribution (black line). The mean and the standard deviation ($\mu$ and $\sigma$) along with expected values for water ($nc_{B,water}$ and $n^2\eta_{B,water}$) are provided. Standard deviation was taken as a measure of spectral resolution. **e**, Anti-Stokes side of the Brillouin spectrum along with Brillouin emission images of a PMMA–water interface, acquired at selected peak frequencies. The interface exhibits sharp contrast in the filtered intensity profiles, which were used to estimate spatial resolution. These profiles were fitted with a Boltzmann sigmoid function to determine the full width at half maximum (FWHM) of the transition zone. Using a 10× objective for both illumination and detection, the spatial resolution was estimated to be ~3 μm from spectral images and ~7 μm from Brillouin amplitude maps derived via DHO fitting (see Supplementary Section 5).

The exceptional spectral resolution and contrast of TFPIs come at a cost. Originally designed to transmit only on-axis light in single-point spectroscopy, the instrument face challenges when used in full-field imaging, where off-axis rays must also be transmitted. Moreover, as part of a micro-spectroscopy system, the TFPI receives light collected through a microscope objective with finite, and often substantial, numerical aperture (NA), necessary for adequate spatial resolution. Consequently, each pixel in the image correspond to the light scattered over a range of angles determined by the finite NA of the detection objective. This results in Brillouin profile broadening which becomes more substantial for low scattering angles [17,30,31]. Furthermore, light from each image location enters the interferometer as a collimated beam at a specific angle relative to its optical axis. The long internal optical paths and the presence of multiple spatial apertures inside TFPI allow it to transmits only a narrow angular subset of these rays, and this filtering varies across the field of view. As a result, the recorded Brillouin spectrum at each pixel is an angle-weighted average, leading to spatially non-uniform spectral parameters, even in homogeneous samples (see Supplementary Section 4).

These distortions are evident in damped harmonic oscillator (DHO) fits applied to bulk water data. As shown in Fig. 1b, the maps of DHO parameters reveal several characteristic artifacts. Field-dependent intensity vignetting in the amplitude map ($A_B$) reflects angular restrictions that limit the effective field of view. Spatial inhomogeneities in the linewidth ($g_B$) indicate non-uniform spectral broadening, likely caused by both, the width of collected scattering angle distribution, and by FPI phase errors from imperfections in mirror flatness or deviations from parallelism, the FPI cavity defects whose effect is amplified by the multi-pass, detuned (tandem) configuration of the interferometer [22,24,32,33]. A gradual tilt in the Brillouin shift ($f_B$) arises from directional filtering of the scattering angle distribution across the field. Finally, a parabolic variation in the offset frequency ($f_0$) results from the effective mirror separation depending on the angle at which light incidence FPI (see Supplementary Section 4).

## 2.2 Modeling and Correction of Angular Artifacts in Brillouin Spectra.

To correct for these effects, we employed a calibration procedure using water as a standard, leveraging its well-characterized speed of sound $c_B$, longitudinal viscosity $\eta_B$, and refractive index $n$. Assuming that no structural relaxation occurs, so $c_B$ and $\eta_B$ remain independent of the acoustic wave vector $q$ [34], any deviation in the measured Brillouin peak shape can be attributed to changes in the scattering angle θ, which alters the magnitude of $q$ (see Supplementary Section 1):

$$f_B(\theta) = \frac{1}{2\pi} n c_B q_n(\theta) \tag{1}$$

$$g_B(\theta) = \frac{1}{4\pi} n^2 \eta_B q_n^2(\theta) \tag{2}$$

To remove explicit dependence on the refractive index, often unknown or spatially varying, we introduce a reduced wave vector $q_n$

$$q_n = \frac{4\pi}{\lambda} \sin(\theta / 2) \tag{3}$$

This formulation allows the Brillouin spectrum to be described in terms of reduced mechanical parameters: the reduced speed of sound, $nc_B$, and the reduced longitudinal viscosity, $n^2\eta_B$, representing the product of the refractive index (or its square) with the underlying mechanical property. The single-scattering-angle DHO spectrum, $I_{DHO}(f,\theta)$, captures how both the Brillouin shift and linewidth vary with scattering angle

$$I_{DHO}(f,\theta) = A_B \frac{2 f_B^2(\theta) g_B(\theta)}{\left((f - f_0)^2 - f_B^2(\theta)\right)^2 + \left(2 f_B(\theta) g_B(\theta)\right)^2} \tag{4}$$

To account for finite scattering angle distribution, each measured spectrum, $I(f)$, was modeled as an angularly averaged sum of single-angle DHO responses, integrated over a scattering angles distribution $p(\theta)$ specific to each pixel – a common strategy for addressing wave-vector indeterminacy arising from collection-angle spread or multiple scattering events [15,17,35,36].

$$I(f) = \int_0^{\pi} I_{DHO}(f,\theta) p(\theta) d\theta \tag{5}$$

A Gaussian form was chosen for $p(\theta)$ as an ad hoc approximation (similarly to [15]), motivated by simplicity and the absence of precise knowledge about the true distribution shape.

This Gaussian $p(\theta)$, characterized by its mean angle $\theta_m$ and standard deviation $\sigma_\theta$, reflects the actual spread of scattering vectors collected by the finite numerical aperture of the microscope objective and filtered by the internal apertures of the FPI, resulting in pixel-dependent distortions in Brillouin line shape. Although ideally treated separately, Brillouin line broadening due to angular averaging and FPI mirror misalignment were both effectively captured by the parameter $\sigma_\theta$.

By fitting experimental spectra from water to eq.(5), with fixed $nc_B$ and $n^2\eta_B$, we estimated the angular distribution parameters $\theta_m$ and $\sigma_\theta$ across the field of view (Fig.1c and Supplementary Section 4). The spatial distribution of $p(\theta)$ is determined by the optical geometry of the system, specifically the interferometer design and the illumination and collection configuration. It is independent of the sample and remains fixed for a given combination of objectives and scattering geometry. Once determined through calibration, the angular distribution can be used to fit spectra from unknown samples, enabling reliable extraction of mechanical parameters free from angular distortion.

### 2.3 Imaging performance on diverse samples.

The resulting corrected maps (Fig. 1d) show high spatial uniformity in $nc_B$, confirming reliable retrieval of Brillouin shift-based metrics. In contrast, $n^2\eta_B$ exhibits a ~40% spread across the field, indicating significantly lower robustness of linewidth-derived quantities. Linewidth is inherently more difficult to estimate with precision, and considerable variability in measured values is commonly observed across different spectrometer systems [4]. These residual uncertainties suggest that viscous property mapping remains challenging when spectral and angular distributions are not tightly controlled. Contributing factors likely include not only angle-averaging effects from the finite numerical aperture, but also instrument-specific imperfections such as inadequacies in FPI mirrors parallelism and non-repeatable alignment during active instrument stabilization routine. We also note that the spectral precision (i.e., the uncertainty in estimating Brillouin shift and linewidth) is higher than the final acoustic precision, indicating that the correction algorithm introduces additional errors in the derived acoustic parameters (Supplementary Section 3).

In addition to imaging mode, the system supports a conventional spectral acquisition mode (see Supplementary Section 2). In this mode, the TFPI is scanned continuously across its full free spectral range (FSR) while a photodiode collects light from the entire field of view, enabling acquisition of a spatially integrated Brillouin spectrum within seconds. This provides a rapid overview of the sample's mechanical heterogeneity prior to full-field imaging. An example is shown in Fig. 1e for a PMMA–water interface, where the integrated anti-Stokes spectrum reveals two distinct Brillouin peaks, corresponding to the different mechanical properties of the two materials. By tuning the interferometer near the peak frequencies of each component, Brillouin emission images specific to PMMA and water were acquired. The presence of two distinct materials and the sharp boundary between them is clearly visible in each single-frequency emission image. These images were used to estimate lateral resolution based on the intensity transition across the material interface, yielding a value of ~3 μm. When the full Brillouin spectrum at each pixel was reconstructed and fitted using the damped harmonic oscillator (DHO) model, the resulting amplitude maps produced a broader transition, corresponding to a spatial resolution of ~7 μm (Supplementary Section 5). The reduction in spatial resolution compared to the theoretical limit set by the objective's numerical aperture can be attributed to the relatively large light-sheet thickness (12 μm) and known limitations in imaging performance inherent to FPI-based systems [25,37].

To demonstrate the utility of the system across biologically and physically relevant scenarios, we evaluated performance on a diverse set of specimens with varying optical properties, structural organization, and mechanical contrast (Fig. 2). A cross-section of Felis catus hair, a hierarchically organized and optically opaque material, was used to test both spatial and mechanical resolution. The integrated Brillouin spectrum (Fig. 2a) revealed three distinct peaks at approximately 5, 7, and 12 GHz, corresponding to water, the keratinous cortex, and the dense cuticle layers, respectively. By tuning the FPI to each peak frequency, Brillouin emission images were generated that clearly delineated these structural domains, consistent with known hair morphology [38] (Fig. 2a–e). Notably, no elastic background leakage was observed in acquired images despite the sample's strong scattering, underscoring the high spectral contrast and selectivity of the TFPI system. Dextran microspheres suspended in water were employed to assess the system's sensitivity to subtle mechanical differences (Fig. 2f). Although the acoustic contrast between the microspheres and the surrounding medium is minimal, the spheres were distinctly resolved in maps of the reduced speed of sound obtained through angular deconvolution approach. To explore biological relevance, we imaged Nicotiana tabacum BY-2 plant cells as a model for plant biomechanics (Fig. 2g). Without the need for labeling or staining, distinct mechanical domains corresponding to the cytoplasm, vacuole, and nucleus were resolved in the reduced speed of sound maps, consistent with expected differences in composition and structure [39]. Finally, we evaluated system performance in optically challenging samples by imaging a section of optically dense onion bulb parenchyma tissue (Fig. 2h). Despite strong light scattering, mechanical structures were clearly visualized, and spatial resolution improved further upon switching from a 5× to a 10× objective. We note that the differences between the 5× and 10× images in Fig. 2h arise from slight axial mismatch and differences in depth of focus, with the 5× objective capturing signal from a larger volume, including surrounding water [17,39]. This can lead to partial spectral averaging, lowering the apparent speed of sound in some regions. These results highlight both the robustness of the system under demanding imaging conditions and flexibility of imaging system.

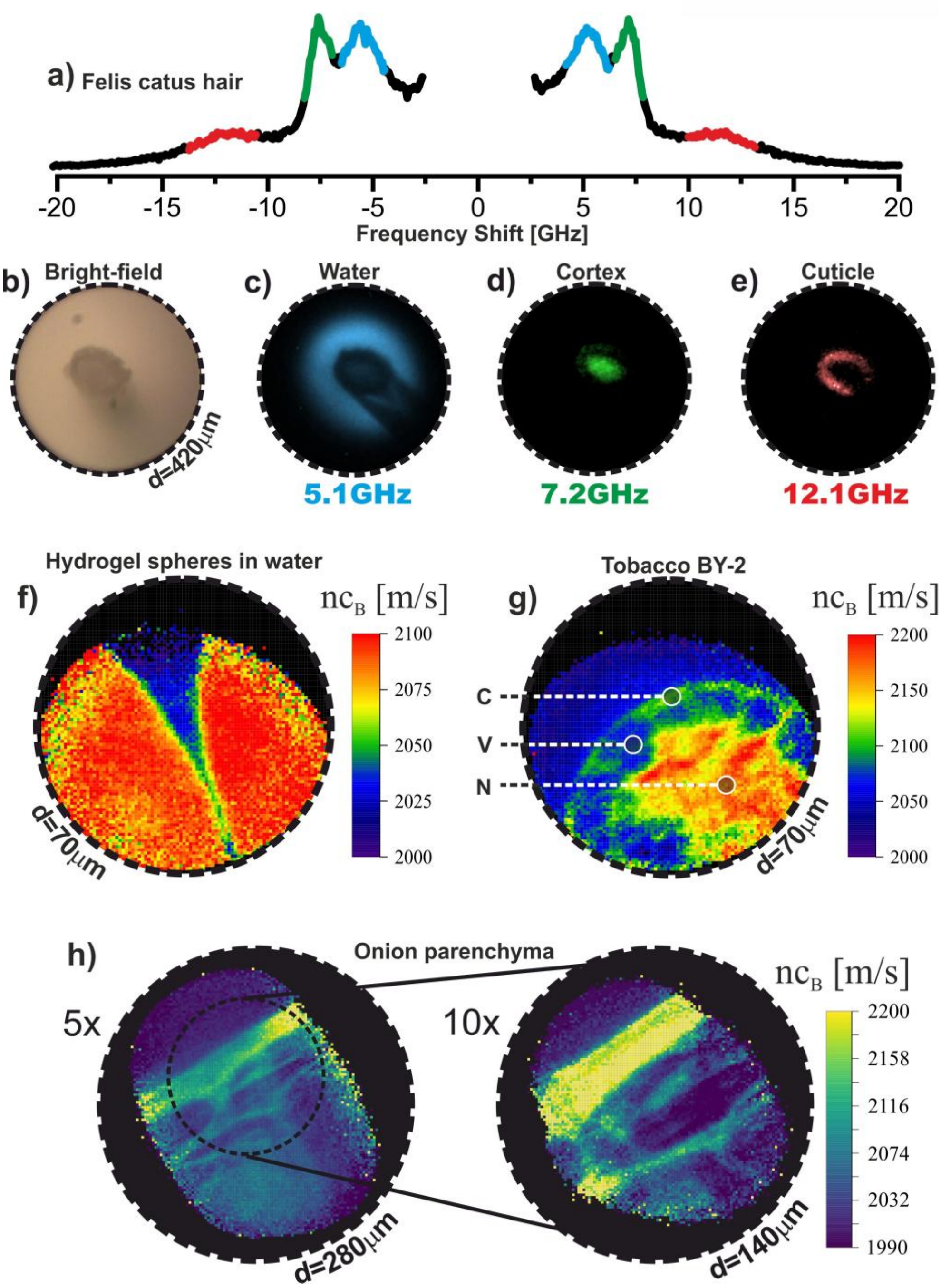

**Fig. 2. Full-field Brillouin imaging of diverse samples. a-e**, Felis catus hair. The integrated Brillouin spectrum (**a**) reveals three distinct peaks at ~5, 7, and 12 GHz, corresponding to water, cortex, and cuticle regions, respectively. The accompanying bright-field image (**b**) and Brillouin emission maps at each peak frequency (**c-e**) demonstrate frequency-specific mechanical contrast and reveal structural features consistent with known hair morphology. **f**, Reduced speed of sound map of dextran hydrogel microspheres suspended in water, illustrating the system's sensitivity to subtle mechanical differences despite minimal refractive index contrast. **g**, Reduced speed of sound map of Nicotiana tabacum BY-2 plant cells showing subcellular mechanical compartmentalization without the need for staining, highlighting potential for label-free mechanobiological analysis. Symbols C, V and N indicate regions of cytoplasm, vacuole and nucleous (with potential reticulum and cytoplasm)

respectively, **h**, Reduced speed of sound maps of onion bulb parenchyma—an optically dense, scattering tissue—acquired under 5× and 10× magnification. Clear mechanical contrast is maintained across both magnifications, demonstrating robustness under demanding imaging conditions. The dashed circles mark the field of view, limited by the physical aperture of the TFPI entrance pinhole.

Operationally, individual spectral images, each acquired at a discrete frequency shift, covered a ~100–200-pixel-diameter field of view depending on camera binning (2 or 1×) and were captured in 0.5–2 seconds. This frame rate enables near real-time visualization of mechanically contrasted Brillouin emission images (Supplementary Video). Complete spectral stacks comprising ~30–40 frequency channels were typically acquired in 30–80 seconds, translating to pixel dwell times in the single-millisecond range, orders of magnitude faster than conventional single-point scanning with FPIs and on par with leading spontaneous Brillouin imaging techniques. Importantly, these acquisition speeds were achieved under low-power illumination conditions, with power densities of only ~0.01 $\mu W/\mu m^3$ (for 50mm lens) or ~0.5 $\mu W/\mu m^3$ (for 10× objective), orders of magnitude lower then when using traditional confocal geometry, and still providing decent spectral precision (15 – 50MHz) (Supplementary section 3 and Table S1).

## 3. Discussion

This work demonstrates that scanning Fabry–Perot interferometers (FPIs), long considered too slow for imaging, can in fact support high-speed, high-contrast full-field Brillouin microscopy when operated in spectral filtering mode. By integrating a canonical multi-pass FPI with light-sheet illumination, we enable spatially resolved mechanical imaging without the need for spatial scanning. A key advantage of this configuration is its ability to acquire Brillouin emission images at selected frequency shifts, an imaging mode unique among spontaneous Brillouin techniques. This feature is particularly beneficial for investigating mechanically heterogeneous samples, where frequency-selective contrast allows targeted visualization of specific material components. Such capability has previously been demonstrated only in stimulated Brillouin gain microscopy [5], where frequency tuning is possible but acquisition is limited to single-point scanning. VIPA-based systems operate in line-scanning mode and rely on spectral multiplexing, which precludes isolating individual frequencies. Similarly, Michelson-type interferometers offer full-field acquisition but operate in the Fourier domain, which does not permit direct frequency filtering or generation of Brillouin emission images. In our system, individual Brillouin emission images can be acquired within fractions of a second, enabling near real-time mechanical contrast visualization (Supplementary video). The full stack of spectral images was acquired within one minute, translating to an effective Brillouin spectrum acquisition time per pixel in the single-millisecond range, orders of magnitude faster than traditional point-scanning FPI systems and on par with modern spontaneous Brillouin imaging techniques. The use of light-sheet illumination ensures uniform excitation and optical sectioning while maintaining significantly lower light doses than single-point approaches (Supplementary Table 1). The high spectral contrast inherent to multi-pass FPIs eliminates the need for additional notch filters to suppress elastic background, even in highly scattering or opaque specimens. Furthermore, although not demonstrated here, FPI mirror reflective coatings can be customized for different laser lines, suggesting that the same setup could be adapted for various excitation wavelengths without requiring hardware modifications.

Although the capabilities of the system are considerable, several inherent limitations remain. The field of view is restricted by the physical size of the entrance pinhole and the limited angular acceptance of the interferometer, which preferentially transmits rays close to the optical axis. This restricted field of view directly impacts the overall throughput by limiting the number of spatial positions that can be multiplexed. The angular filtering introduces field-dependent variation in the effective Brillouin shift (see $f_B$ in Fig.1b). Additional constraints stem from the use of a collimated FPI geometry, in which the effective mirror separation depends on the angle of incidence. As a result, the transmitted frequency at a fixed FPI setting varies across the field (see $f_0$ in Fig.1b). Furthermore, minor imperfections in in mirror parallelism, surface flatness, or reflectivity can lead to spatially varying transmission profiles (non-uniform cavity FPIs maps)[24,32,37]. In addition, the internal angular filtering in our current design introduces further complexity. At off-axis positions, the transmitted wave-vector distribution becomes skewed, which could lead to a narrowing of the measured Brillouin profile. Simultaneously, the same filtering mechanism may degrade the spectral resolution (finesse), potentially causing artificial broadening of the Brillouin peaks (Supplementary Section 4). The interplay between these opposing effects remains to be fully understood.

In the current setup, FPI alignment routine [40] performs intermittent adjustment cycles every few seconds, introducing small but non-reproducible changes in mirror alignment. These variations, amplified in the multi-pass detuned (tandem) configuration, lead to subtle but significant field-dependent spectral broadening, particularly affecting linewidth-based measurements. To address this, future systems could benefit from real-time, inline monitoring of interferometer performance, ideally decoupled from the measurement path. While our angular deconvolution reduces Brillouin shift non-uniformity, viscosity estimates remain more sensitive to these distortions. As the accuracy of acoustic parameters lags behind spectral precision, faster and more robust correction algorithms are needed. Performance could also be improved by increasing the collimation f-number and using wider-aperture optics to enhance angular acceptance and field uniformity. Alternatively, adopting a telecentric design [23–25,32] may offer further benefits for dedicated imaging systems.Another limitation of our system is its reliance on light-sheet illumination, which, while providing optical sectioning and low photodose, requires separate illumination and detection paths. This necessitates sample accessibility from both sides, and refractive index mismatches can introduce wave-vector uncertainty. Additionally, the 90° scattering geometry limits the scattering vector, and when using high-NA objectives, can lead to substantial Brillouin linewidth broadening. A backscattering configuration (~180°) would be preferable and could potentially be implemented using axial-plane or spinning-disc confocal microscopy–style illumination schemes [41,42].

Several practical strategies are available to further improve acquisition throughput. One clear option is to increase the illumination power density, while still staying within acceptable photodose limits. In this study, both sides of the Brillouin spectrum (Stokes and anti-Stokes) were acquired to account for the spectral offset $f_0$. Since $f_0$ is determined by the TFPI collimation geometry, it can be treated as a fixed parameter once characterized. Consequently, because the Brillouin spectrum is symmetric about $f_0$ (see Fig.1a and 2a), acquiring only one side of the spectrum can halve the number of required spectral images without loss of information. Additionally, the number of frequency points per spectrum side can be reduced by using optimized spectral fitting methods that accurately reconstruct the Brillouin peak with fewer data points. Further gains could be achieved with more sensitive cameras and simplified interferometer optics that preserve the necessary spectral contrast.

Altogether, this study demonstrates that the scanning tandem Fabry–Perot interferometer can be effectively repurposed for high-contrast, full-field Brillouin imaging. The instrument, already widespread in many laboratories, particularly in non-biological fields, offers a practical foundation for extending Brillouin methods beyond traditional point-scanning approaches. While the system excels in applications requiring extreme spectral resolution and contrast, it also offers flexibility: future implementations may sacrifice some of its contrast in favor of simpler, brighter designs optimized for specific imaging needs. We provide experimental validation repurposing a commercial TFPI and identify key spectral artifacts that limit performance in its current form. We also suggest practical strategies for mitigating these limitations. Given the widespread availability of TFPIs, we anticipate that our approach will enable rapid, high-throughput Brillouin-based imaging in a broad range of applications, both within and beyond the life sciences.

## 4. Methods

**4.1 Optical setup for full-field Brillouin imaging.** Full-field Brillouin imaging was implemented by adapting a commercial Sandercock-type Tandem Fabry–Perot interferometer (The Table Stable Ltd., TFPI-2). The system combines static light-sheet illumination, orthogonal scattering collection, and sequential frequency filtering via a scanning interferometer (Supplementary Fig. S2a). The excitation source was a CW laser (532 nm, 250 mW, 2 mm diameter). A cylindrical lens (f = 200 mm) focused the beam onto the back focal plane of the illumination objective (either 50 mm achromatic doublet or Olympus 10×), producing a light sheet. For the 50 mm lens, the resulting light-sheet had a theoretical width of ~500 μm, Rayleigh range of ~1600 μm, and beam waist (thickness) of ~32 μm, yielding an illumination power density of ~0.01 $\mu W/\mu m^3$. When the 10× objective was used, the light-sheet dimensions were ~180 μm in width, ~210 μm in Rayleigh range, and ~12 μm in beam waist, with a corresponding power density of ~0.5 $\mu W/\mu m^3$. Orthogonally scattered light was collected by detection objectives (5×, 10×, or 20×) and re-imaged onto the entrance aperture (1 mm diameter) of the TFPI using a tube lens (f = 150 mm). An integrated CMOS camera allowed bright-field inspection. The light from TFPI entrance aperture was collimated and passed through two FP cavities in a six-pass configuration. Spectrally filtered light was directed via a flip mirror to either a CCD camera (FLIR Grasshopper3) or a photodiode, depending on the acquisition mode. All six interferometer axes (four mirror tilts, differential mirror spacing and the common mirror separation), optical shutters and flip mirror were controlled via NI LabVIEW software and NI DAQ boards. The FP mirrors separation was fixed at 3 mm, yielding a 50 GHz free spectral range (FSR) as a compromise between spectral resolution and field of view.

**4.2 Acquisition workflow.** Measurements proceeded through three operating modes: stabilization, spectral, and imaging. In stabilization mode, a reference beam (extracted from incident laser beam) was routed through the TFPI to a photodiode. A custom optimization routine adjusted all six TFPI degrees of freedom to maximize transmission and lock the transmission frequency to the reference laser line. This process was completed in ~1.5 s and repeated every ~10–15 s for maintaining instrument stability (Supplementary Fig. S2b). In spectral mode, light scattered from the full field of view was passed to the photodiode, and mirror spacing was scanned across the FSR to obtain a spatially integrated Brillouin spectrum. In imaging mode, light was redirected to the CCD camera. The set of frequency shifts (mirror separations) was selected based on the outcome of preliminary spectral measurements. At each discrete frequency shift, a spectral image was recorded. A complete Brillouin map typically consisted of ~40 frames (20 Stokes and 20 anti-Stokes), acquired over 30–80 s depending on the camera pixel binning.

**4.3 Spectral, Acoustic and Spatial Resolution Estimation.**

Spectral precision was evaluated using two complementary approaches: per-pixel and field-of-view (FOV)–averaged analysis. In both cases, the DHO model was fitted pixel-by-pixel across seven independent measurements of a homogeneous water sample. For per-pixel precision, the standard deviation of the extracted Brillouin shift ($f_B$) and linewidth ($g_B$) was computed at each pixel location across the repeated measurements, resulting in spatial maps of precision for both parameters. FOV-averaged precision was estimated by analyzing the distribution of $f_B$ and $g_B$ values across the entire field of view within each measurement. The spread of these distributions, calculated over multiple acquisitions, served as a global estimate of the system's spectral stability. This dual approach captures both local variations in measurement repeatability and overall system precision across the imaging area (Supplementary Section 3).

Acoustic resolution was assessed using a two-step angular deconvolution approach. In the first step, a homogeneous calibration sample (pure water) with known mechanical properties (reduced speed of sound $nc_B$ and reduced longitudinal viscosity $n^2\eta_B$) was used to determine pixel-wise angular distribution parameters $\theta_m(x,y)$ and $\sigma_\theta(x,y)$. These were obtained by fitting the measured spectra to a model based on Gaussian-distributed scattering angles. With angular maps fixed, the second step involved analyzing spectra from a second water dataset, fitting for $nc_B$ and $n^2\eta_B$ while keeping maps of $\theta_m$ and $\sigma_\theta$ constant. This yielded spatially resolved maps of reduced acoustic and viscous parameters. Histograms of the retrieved mechanical parameters were compiled from ten independent measurements. The averaged histogram for each parameter was fitted with a normal distribution, and the standard deviation was taken as a measure of spectral resolution (Supplementary Section 3).

Spatial resolution was assessed using a two-material sample composed of PMMA and water, chosen for their distinct mechanical properties and well-defined boundary. In the first approach, spectral images were acquired at the Brillouin peak frequencies of each material, and intensity profiles across the interface were extracted and fitted with a Boltzmann sigmoid function to determine the full width at half maximum (FWHM), yielding an estimated resolution of approximately 3 μm. In the second approach, a full stack of spectral images was recorded to reconstruct the Brillouin spectrum at each pixel. The resulting spectra were fitted with a sum of two damped harmonic oscillator (DHO) functions, one for each material, and amplitude maps were generated. Line profiles across the boundary in these maps were again fitted with Boltzmann functions, giving a broader FWHM of approximately 7 μm. Together, these complementary methods demonstrate a spatial resolution in the range of 3–7 μm, depending on the analysis approach (Supplementary Section 5).

**4.4 Samples.** White *Felis catus* hair was collected and cross-sectioned with scissors. The hair was mounted onto a syringe needle attached to an XYZ translation stage for controlled movement and submerged in distilled water. The sample was oriented so that the light sheet illuminated the cross-sectioned surface. A suspension of *Nicotiana tabacum* BY-2 cells was poured into a 5×5 mm quartz cuvette, and Brillouin signal was collected from the illuminated region near the cuvette corner (Details on the BY-2 cell culture in [39]). Dextran microspheres (Sephedex, diameter ~400 μm) in dry powder form were suspended in distilled water. After swelling into hydrogel form, the sample was transferred to a 5×5 mm quartz cuvette, and measurements were conducted similarly to the BY-2 cells. A ~2×2×2 mm piece of onion bulb parenchyma was cut and mounted onto a syringe needle attached to an XYZ stage. The specimen was submerged in distilled water, and imaging was performed near the water–onion boundary.

## Funding
National Science Centre of Poland OPUS grants: 2021/41/B/ST5/03038 and 2023/51/B/ST3/01995

## Acknowledgements
Author gratefully acknowledge dr Tomasz Skrzypczak for supplying the *Nicotiana tabacum* BY-2 samples and prof. Krzysztof Dobek for valuable discussions.

## Disclosures
The authors declare no conflicts of interest.

## Data availability
Data underlying the results presented in this paper may be obtained from the corresponding author upon reasonable request.